\documentclass{aa}

\usepackage{graphicx}
\usepackage{amsmath}
\usepackage{txfonts}
\usepackage{natbib}
\usepackage{caption}
\usepackage{amssymb}
\usepackage{hyperref}
\usepackage{pdflscape}
\usepackage{orcidlink}
\usepackage{soul}
\usepackage{threeparttable}
\usepackage{multirow}
\usepackage{enumerate}
\usepackage{indentfirst}
\usepackage{stackengine}
\newcommand\xrowht[2][0]{\addstackgap[.5\dimexpr#2\relax]{\vphantom{#1}}}
\usepackage{subcaption}

\hypersetup{colorlinks=true, linkcolor=red, citecolor=blue, urlcolor=blue, bookmarks=true}

\def\gsimeq{\hbox{\raise0.5ex\hbox{$>\lower1.06ex\hbox{$\kern-1.07em{\sim}$}$}}}
\def\lsimeq{\hbox{\raise0.5ex\hbox{$<\lower1.06ex\hbox{$\kern-1.07em{\sim}$}$}}}

\bibpunct{(}{)}{;}{a}{}{,}

\begin{document}

\title{Active galactic nuclei-heated dust revealed in ``little red dots''}

   \subtitle{}

   \author{I.~Delvecchio
          \inst{1}\fnmsep
          \thanks{ivan.delvecchio@inaf.it} \orcidlink{0000-0001-8706-2252}
          \and E.~Daddi \inst{2} \orcidlink{0000-0002-3331-9590}
          \and B.~Magnelli \inst{2} \orcidlink{0000-0002-6777-6490}
          \and D.~Elbaz \inst{2} \orcidlink{0000-0002-7631-647X}
          \and M.~Giavalisco \inst{3} \orcidlink{0000-0002-7831-8751}
          \and A.~Traina \inst{1} \orcidlink{0000-0003-1006-924X}
          \and G.~Lanzuisi \inst{1} \orcidlink{0000-0001-9094-0984}
          \and H.B.~Akins \inst{4} \orcidlink{0000-0003-3596-8794}
          \and S.~Belli \inst{5} \orcidlink{0000-0002-5615-6018}
          \and C.M.~Casey \inst{6} \orcidlink{0000-0002-0930-6466}
          \and F.~Gentile \inst{2,1} \orcidlink{0000-0002-8008-9871}
          \and C.~Gruppioni \inst{1} \orcidlink{0000-0002-5836-4056}
          \and F.~Pozzi \inst{5} \orcidlink{0000-0002-7412-647X}
          \and G.~Zamorani \inst{1} \orcidlink{0000-0002-2318-301X}
          }

\authorrunning{I. Delvecchio et al.}
   \institute{ INAF – Osservatorio di Astrofisica e Scienza dello Spazio di Bologna, Via Gobetti 93/3, I-40129 Bologna, Italy
\and Universit\'e Paris-Saclay, Universit\'e Paris Cit\'e, CEA, CNRS, AIM, 91191, Gif-sur-Yvette, France
\and Astronomy Department, University of Massachusetts, Amherst, MA 01003, USA
\and The University of Texas at Austin, 2515 Speedway Boulevard Stop C1400, Austin, TX 78712, USA
\and University of Bologna, Department of Physics and Astronomy (DIFA), Via Gobetti 93/2, I-40129, Bologna, Italy
\and Department of Physics, University of California, Santa Barbara, Santa Barbara, CA 93106, USA
}

   \date{}

  \abstract
  {Little red dots (LRDs) are a puzzling population of extragalactic sources whose origin is highly debated. In this {work}, we performed a comprehensive stacking analysis of NIRCam, MIRI, and ALMA images of a large and homogeneously selected sample of LRDs from multiple JWST Legacy fields. We report clear evidence of hot-dust emission in the median stacked spectral energy distribution (SED) that features a rising near-infrared continuum up to rest-frame $\lambda_{\rm rest}$$\sim$~3$\mu$m, which is best explained by a standard dusty active galactic nucleus (AGN) structure. Although LRDs are likely to be a heterogeneous population, our findings suggest that most ($\gtrsim$50\%) LRDs show AGN-heated dust emission, regardless of whether the optical and ultraviolet (UV) continua are stellar or AGN-dominated. In either case, the best-fit dusty-AGN SED, combined with the lack of X-ray detection in the deep Chandra stacks, suggests that Compton-thick ($N_{\rm H}$$>$3$\times$10$^{24}$~cm$^{-2}$) gas obscuration is common, and likely confined within the dust sublimation radius ($R$$_{\rm sub}$$\sim$0.1~pc). Therefore, we argue that AGN-heated dust does not directly obscure either the optical-UV continuum or the broad-line region emission, in order to explain the observed blue UV slopes and prominent Balmer features. While a gas-dust displacement is in line with several models, the formation scenario (in-situ or ex-situ) of this pre-enriched hot dust remains unclear.}


\vspace{-0.5cm}

\titlerunning{AGN dust in LRDs}
\authorrunning{I. Delvecchio et al.}

  \maketitle

\section{Introduction} \label{intro}

The James Webb Space Telescope (JWST) has ushered extragalactic astronomy into a transformative era, offering unprecedented access to the high-redshift ($z$$>$4) universe with the discovery of previously missed active galactic nuclei (AGN; e.g. \citealt{Matthee+2024}). These AGNs are mostly identified via broad emission lines (full width at half maximum$>$1000--2000~km~s$^{-1}$), corresponding to bolometric luminosities of L$_{\rm bol,AGN}$$\sim$10$^{44-46}$~erg~s$^{-1}$ and harboring supermassive black holes (SMBHs, with mass $M_{\rm BH}$$\sim$$10^6 - 10^8$$~M_\odot$). In contrast to their lower-$z$ analogs, they do not show significant X-ray or radio emission (\citealt{Ananna+2024}, \citealt{Mazzolari+2025}). A non-negligible fraction of these AGN ($\sim$10--30\%; \citealt{Hainline+2025}) feature unique and puzzling properties, and are dubbed ``little red dots'' (LRDs, e.g., \citealt{Matthee+2024}, \citealt{Greene+2024}, \citealt{Labbe+2025}). These LRDs are extremely compact (often unresolved at $\lesssim$100~pc scales) and show red optical colors combined with blue ultraviolet (UV) slopes, which result in a characteristic ``V-shaped'' spectral energy distribution (SED), with a sharp turnover near the Balmer break at rest $3645$~\AA~(e.g., \citealt{Setton+2024}, \citealt{Hainline+2025}). The vast majority of them are seen at 4$<$$z$$<$8, and about $\sim$60\% of the spectroscopically confirmed LRDs display broad emission lines (e.g., \citealt{Greene+2024}; \citealt{Hviding+2025}). Notably, LRDs appear to be up to two orders of magnitude more abundant than luminous quasars at z$>$4 (\citealt{Kokorev+2024}, \citealt{Akins+2025}), which makes them accessible even in narrow-field JWST surveys.

However, an observational consensus on the physical nature of LRDs is still lacking.
On the one hand, if LRDs are compact star-forming galaxies (e.g., \citealt{Labbe+2023}), their stellar masses ($M_{\star}$$\sim$10$^{8-11}$~M$_{\odot}$) would imply extreme stellar densities \citep{Pacucci+2024}. On the other hand, the red optical colors and broad emission lines hint at an AGN origin (\citealt{Inayoshi+2020}, \citealt{Volonteri+2021}, \citealt{Ji+2025}), from which the (e.g., H$_{\alpha}$-based) $M_{\rm BH}$ estimates would imply overmassive black holes (BHs) ($M_{\rm BH}$/$M_{\star}$$\gtrsim$1:100; see e.g., \citealt{Pacucci+2023}; \citealt{Juodzbalis+2024}; \citealt{Maiolino+2024}; \citealt{Juodzbalis+2025}) compared to local scaling relations \citep{Reines+2015}, even without accounting for dust attenuation (e.g., \citealt{Greene+2025}).

In the AGN scenario, the optical and/or UV continuum would be moderately reddened by dust ($A_V$$\sim$$2$--$3$ mag; \citealt{Casey+2024}; \citealt{Akins+2025}; \citealt{Wang+2025}). However, this interpretation is apparently at odds with their extreme X-ray and radio weakness implied by the lack of stacked X-ray (\citealt{Ananna+2024}; \citealt{Yue+2024}; \citealt{Maiolino+2025}) and radio emission (\citealt{Akins+2025}; \citealt{Gloudemans+2025}). Recent studies attribute the X-ray weakness to super-Eddington accretion (\citealt{Madau+2024}; \citealt{Lambrides+2024}; \citealt{Pacucci+2024}; \citealt{King2025}; \citealt{Madau2025}; but see e.g., \citealt{Sacchi+2025}).

To also alleviate the tension between the prominent optical and/or UV Balmer features and the apparent X-ray weakness, several models postulate the presence of a turbulent, dust-free, and highly dense gas envelope (e.g., \citealt{Liu+2025}; \citealt{Cenci+2025}; \citealt{Inayoshi+2025}; \citealt{Naidu+2025}; \citealt{Rusakov+2025}; \citealt{deGraaff+2025}). These conditions explain the strong Balmer breaks and the red continua through pure gas opacity, since the gas temperature is too high for the dust to survive. In addition, the broad Balmer lines and other absorption features can be explained via electron scattering (e.g., \citealt{Rusakov+2025}; \citealt{Inayoshi+2025b}) or resonant Balmer scattering (e.g., \citealt{Naidu+2025}). These models mostly differ on the nature of the central object within the gas envelope, which could be either a direct-collapse BH (e.g., \citealt{Cenci+2025}), a Super-Eddington BH (\citealt{Liu+2025}), a late-stage ``quasi star'' \citep{Begelman+2025}, a ``BH-Star'' (e.g. \citealt{Naidu+2025}), or a metal-free object such as a super-massive star \citep{Nandal+2025}.

Such a dust paucity is further supported by the non-detection from deep Atacama Large Millimeter Array (ALMA) stacking, which poses a stringent upper limit to the cold dust budget, corresponding to M$_{\rm dust}$$<$10$^{6}$~M$_{\odot}$ at $z$$\sim$6 (\citealt{Casey+2025}; see also \citealt{Chen+2025}; \citealt{Setton+2025}). In the rest-frame near and mid-IR, a broad diversity of LRD SEDs has been reported in the recent literature. Stacking of Mid-Infrared Instrument (MIRI)-undetected LRDs reveals a flat SED beyond rest 1.6~$\mu$m (e.g., \citealt{Williams+2024}; \citealt{Carranza-Escudero+2025}), in contrast to the rising mid-infrared (MIR) continuum expected in obscured AGN, which conveys the idea that MIR emission in LRDs is typically dominated by dust-poor stellar populations and not AGN dust. However, based on individual MIRI detections, other studies found that some LRDs display a flat or rising MIR continuum (\citealt{Perez-Gonzalez+2024}; \citealt{Barro+2024}; \citealt{Leung+2025}; \citealt{deGraaff+2025}; \citealt{Ronayne+2025}), which is best fitted via composite stellar and AGN templates, suggesting a heterogeneous origin of the IR emission of LRDs.

In this highly debated framework, we caution that previous MIRI-SED studies of LRDs might have been biased toward bright individual objects (e.g., \citealt{Labbe+2024}; \citealt{Barro+2024}; \citealt{Chen+2025}; \citealt{deGraaff+2025}; \citealt{Wang+2025}), or possibly affected either by individual MIRI non-detections (e.g., \citealt{Perez-Gonzalez+2024}) or by stacking peculiar LRD subsamples (e.g., $H$-dropouts; \citealt{Williams+2024}), whose SED could intrinsically differ from that of ``typical'' LRDs. To explore this possibility, in this {paper} we present a comprehensive stacking analysis of a large and homogeneously selected LRD sample with broad Near Infrared Camera (NIRCam), MIRI, and ALMA coverage. Throughout, we assume a \citet{Chabrier2003} initial mass function (IMF) and a flat $\Lambda$CDM cosmology with $H_0$ = 70~km/s/Mpc, $\Omega_m = 0.30$, and $\Omega_{\Lambda} = 0.70$.

\begin{table*}[!t]
\caption{List of the survey fields considered in this study.}
\centering
\begin{tabular}{c|cccc}
\hline\hline
Survey/Field   &  Instrument & Filters                        &   Area                   &  $N$$_{\rm stack}$ range  \\
               &             &                                &  (arcmin$^2$)            &     (spec)        \\
\hline
CEERS/EGS      &    NIRCam   & [f090w, f115w, f150w, f200w, f277w, f356w, f444w]             & 88.6  & 63 (31) \\
               &    MIRI     & [f560w, f770w, f1000w, f1280w,                               &       &         \\
               &             & f1500w, f1800w, f2100w]                                      & 55.0  & 2--39 (2--19)  \\
\hline
JADES/GOODSS  &    NIRCam   & [f090w, f115w, f150w, f200w, f277w, f356w, f444w]              & 62.1  & 44 (8) \\
SMILES/GOODSS &    MIRI     & [f560w, f770w, f1000w, f1280w,                                &       &         \\
              &             & f1500w, f1800w, f2100w, f2550w]                               & 34.0  &  22 (2) \\
A3-GOODSS     &    ALMA     & Band-6 (1080$\mu$m)                                          & 86.8  & 34 (4)  \\
\hline
PRIMER/COSMOS  &    NIRCam   & [f090w, f115w, f150w, f200w, f277w, f356w, f444w]             & 138.9 & 79 (15) \\
               &    MIRI     & [f770w, f1800w]                                               & 73.0  & 41 (7) \\
A3-COSMOS      &    ALMA     & Band-6 (1080$\mu$m)                                         & 127.9 & 36 (3)  \\
\hline
PRIMER/UDS     &    NIRCam   & [f090w, f115w, f150w, f200w, f277w, f356w, f444w]             & 243.0 & 116 (42) \\
               &    MIRI     & [f770w, f1800w]                                               & 130.0 & 61 (23) \\
\hline
{All fields}  &      {NIRCam}   & --                                          & 531.7   &  $\leq$302 ($\leq$96) \\
                    &      {MIRI}     & --                                               & 292.0   &  $\leq$163 ($\leq$51) \\
\hline
\hline

\end{tabular} \label{tab:sample}
\tablefoot{For each survey, the sky area, the available filters, and the corresponding number of LRDs (including in brackets those with a spectroscopic redshift) are reported. The range of $N$$_{\rm stack}$ reflects the varying area across different filters of the same instrument. The global sample is reported at the bottom of this table. Field acronyms: EGS = Extended Groth Strip; GOODS-S = Great Observatories Origins Deep Survey--South; UDS = UKIDSS Ultra Deep Survey; COSMOS = Cosmic Evolution Survey.}
\end{table*}

\section{Photometric selection of LRDs} \label{sample}

We selected LRDs from the compilation of \citet{Kocevski+2025}, which consists of 341 photometrically selected LRDs in the redshift range $z$$\sim$2--11 that were retrieved from several JWST/NIRCam surveys. These LRDs were homogeneously selected to display compact morphologies and ``V-shaped'' continua from spectral slope fitting between blueward and redward of the Balmer break (rest-frame 3645~\AA). These criteria enable a self-consistent search for LRD candidates with red optical and blue UV colors over a wide redshift interval (\citealt{Hainline+2025}), although we acknowledge that it does not guarantee an homogeneous physical selection of LRDs.

\begin{figure}[!h]
\centering
     \includegraphics[width=\linewidth]{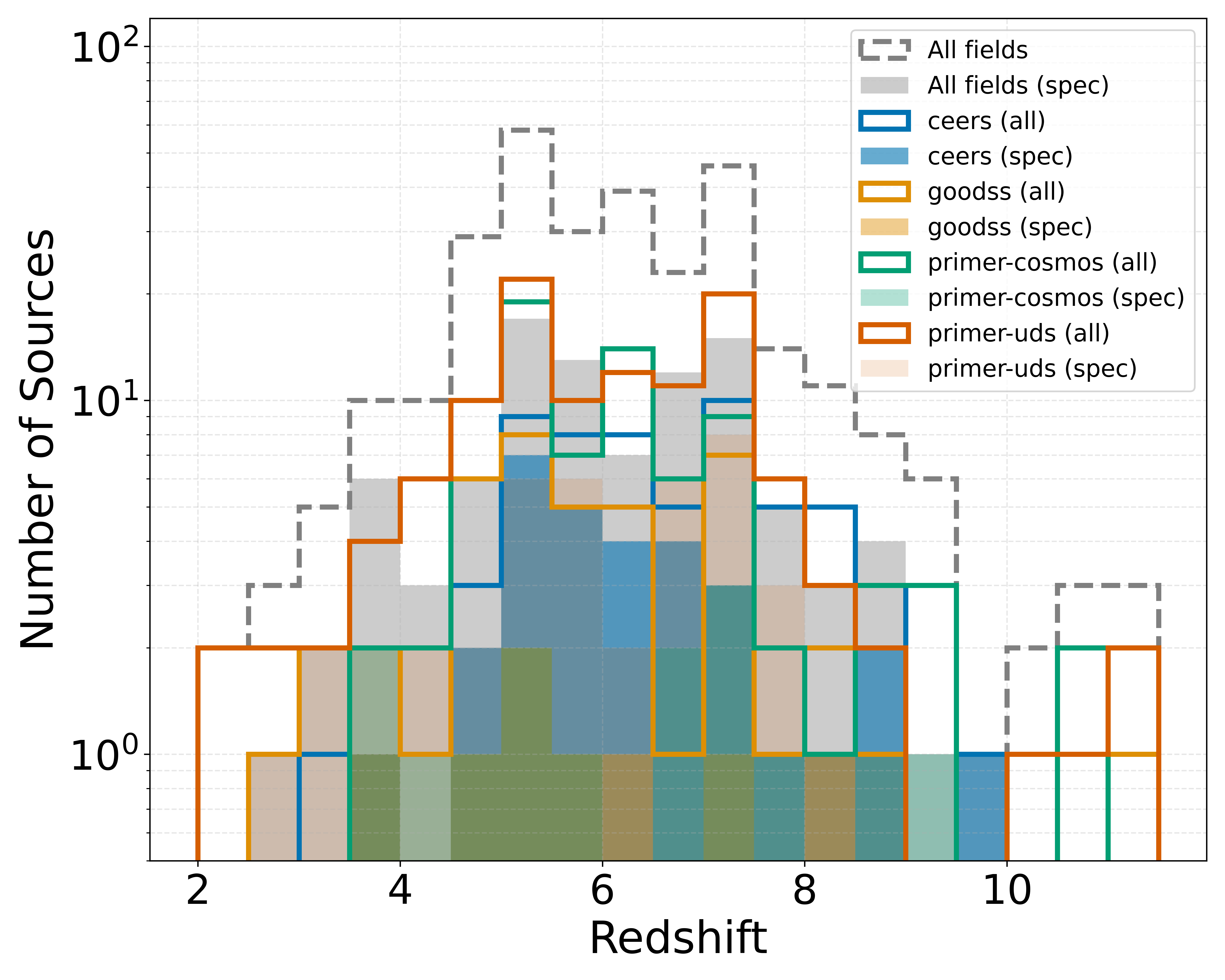}
 \caption{\small Redshift distribution of the final LRD sample (dashed black histogram) used in this work. The full sample is split by field (CEERS, GOODSS, PRIMER-COSMOS, PRIMER-UDS), as highlighted by the colored open histograms, while filled histograms mark the corresponding subset with spectroscopic redshifts.
 }
   \label{fig:z}
\end{figure}

We replaced photometric redshifts ($z_{\rm phot}$) with NIRSpec-based spectroscopic redshifts ($z_{\rm spec}$) taken from the NIRSpec Merged Table (version 4\footnote{\url{https://zenodo.org/records/15472354}}; \citealt{Valentino+2025}) of the public Dawn JWST Archive (DJA; using {\sc msaexp}; \citealt{Brammer2023b}; \citealt{deGraaff+2024}; \citealt{Heintz+2025}), whenever it was of high quality (grade=3, see \href{https://s3.amazonaws.com/msaexp-nirspec/extractions/nirspec_graded_v3.html}{archive}). We augmented this list by adding $z_{\rm spec}$ measurements from the JADES public database\footnote{\url{https://jades.herts.ac.uk/search/}} (for the GOODS-S). This enhanced the spectroscopic fraction from 11\% (39/341) to 31\% (107/341). The mutual agreement between $z_{\rm phot}$ and $z_{\rm spec}$ estimates is very good, with a median absolute deviation (MAD) $\langle$$|z_{\rm spec} - z_{\rm phot}| / (1+z_{\rm spec})$ $\rangle$$\approx$1.6\%, and only five catastrophic outliers with $|z_{\rm spec} - z_{\rm phot}|$/(1+$z_{\rm spec}$)$>$0.2. Hence, we do expect a negligible impact of photo-$z$ uncertainties on the median stacks. Fig.~\ref{fig:z} displays the redshift distribution of our final LRD sample (302 sources, grey dashed), split by field (colored open histograms). Filled histograms highlight the spectroscopic subsets. Corresponding numbers are listed in Table~\ref{tab:sample}.

Finally, we restricted ourselves to the NIRCam fields with (at least partial) MIRI coverage in two or more bands (including 18$\mu$m and/or 21$\mu$m; see Table~\ref{tab:sample}), which yield the strongest IR-SED constraints. Our final sample counts 302 LRDs taken from the following surveys: the Cosmic Evolution Early Release Science Survey (CEERS; \citealt{Finkelstein+2023}) in the EGS; the JWST Advanced Deep Extragalactic Survey (JADES; \citealt{Eisenstein+2023}) in the Great Observatories Origins Deep Survey-South (GOODS-S) field, complemented by MIRI imaging from the Systematic Mid-infrared Instrument Legacy Extragalactic Survey (SMILES; \citealt{Rieke+2024}; \citealt{Alberts+2024}); the Public Release IMaging for Extragalactic Research (PRIMER\footnote{\url{https://primer-jwst.github.io/}}) survey in the COSMOS and UDS fields. Out of 302 LRDs, 163 (54\%) are covered by MIRI 7.7$\mu$m, 126 (42\%) also by MIRI-18$\mu$m, and {96 (32\%) have a robust $z_{\rm spec}$}. Moreover, we exploited archival ALMA/B6 (1.1~mm) data from the combined Automated Mining of the Public ALMA Archive (A3) A3COSMOS+A3GOODSS surveys (\citealt{Liu+2019}; \citealt{Adscheid+2024}; \citealt{Magnelli+2024}), which are available for a total of 70 LRDs in the COSMOS and GOODS-S fields. The number of sources split up by field, photometric coverage, and spectroscopic redshifts is summarized in Table~\ref{tab:sample}.

\begin{figure*}
\centering
     \includegraphics[width=\linewidth]{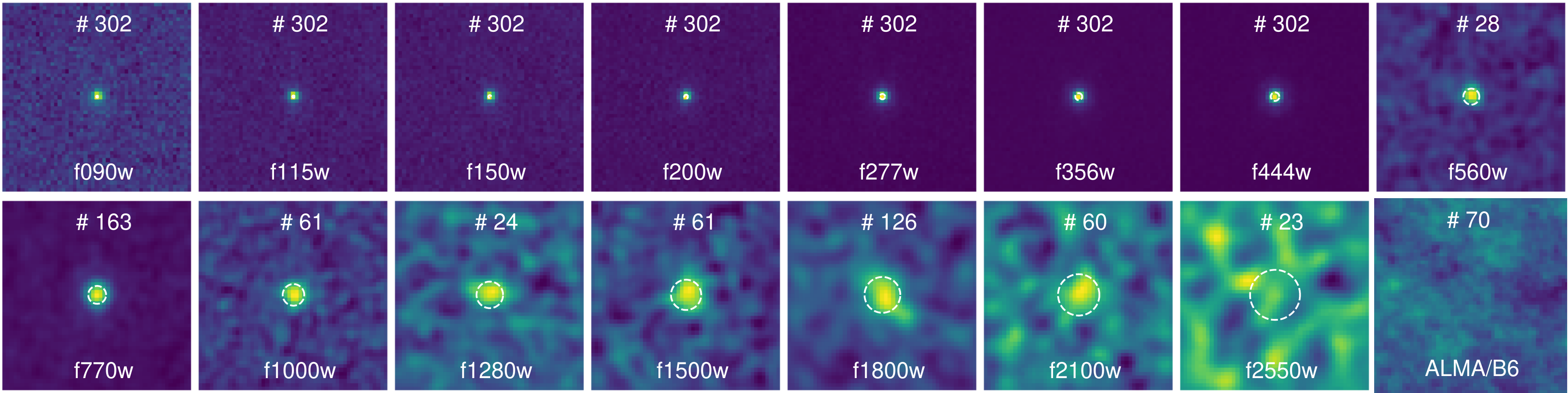}
 \caption{\small Median stacked cutouts (3$^{\prime \prime}$$\times$3$^{\prime \prime}$) of LRDs in NIRCam (7; from f090w to f444w), MIRI (8; from f560w to f2550w), and ALMA/B6 (1.1~mm) bands. All NIRCam and MIRI images have the same pixel size (0.06$^{\prime \prime}$). Each cutout reports the corresponding PSF's FWHM (white circle) and the number of stacked sources. To ease visualization, a smoothing with a Gaussian kernel was applied to all MIRI images, with a larger radius for larger FWHM: two pixels for f560w, f770w, and f1000w; three pixels for f1280w and f1500w; four pixels for f1800w and f2100w, and five pixels for f2550w. The stacked ALMA/B6 image was obtained in the $uv$-plane and imaged at 0.1$^{\prime \prime}$/px scale. Except for MIRI/f2550w and ALMA/B6, all other median stacks lead to a S/N$>$3 detection.
 }
   \label{fig:cutouts}
\end{figure*}

\section{Stacking method and results} \label{stacking}

Here we outline the input images used for stacking and the stacking method adopted in this work. We performed median stacking in all available NIRCam bands (seven in total) and MIRI bands (two in PRIMER, seven in CEERS, and eight in SMILES), at the optical position of all LRDs. We used a fixed pixel scale of 0.06$^{\prime \prime}$, and we extracted fluxes from a circular aperture with fixed radius $r$=0.25$^{\prime \prime}$. Image and noise maps were retrieved from the corresponding mosaics of each field (see below). At this step, we discarded narrow or medium-band filters (whenever available) to mitigate potential flux boosting in the case of strong emission lines.

We note that stacking was performed in the observed frame, since stacking in the rest-frame would require us to K-correct fluxes by assuming an intrinsic SED, which is instead what we would like to assess, particularly in the rest-frame near-IR (NIR, 1--3$\mu$m), where an AGN-like or a galaxy-like SED alters the interpretation. However, a rest-frame stacking analysis was carried out in \citet{Casey+2024} that delivered consistent results with ours. We discuss this comparison in Sect.~\ref{sed_comparison}.

\subsection{Input image maps}

We retrieved PRIMER-COSMOS images as part of the COSMOS-Web team, where the full PRIMER-COSMOS mosaic has been reduced following the same procedure as with the COSMOS-Web data (\citealt{Franco+2025}; \citealt{Harish+2025}). All NIRCam and MIRI mosaics have a pixel size of 0.06$^{\prime \prime}$.

For the GOODS-S, we retrieved NIRCam and MIRI images from the JADES and SMILES surveys publicly available on the MAST\footnote{SMILES: \url{https://archive.stsci.edu/hlsp/smiles}; JADES: \url{https://archive.stsci.edu/hlsp/jades}} archive. While SMILES maps are already at 0.06$^{\prime \prime}$/px, JADES maps were rescaled from their native resolution (0.03$^{\prime \prime}$/px) to 0.06$^{\prime \prime}$/px for consistency with the other fields..

For CEERS and PRIMER-UDS, which are unavailable on MAST, we leveraged the DJA (\citealt{Valentino+2023}; using {\sc grizli}, \citealt{Brammer2023a}), where the full reduced mosaics are publicly available in each NIRCam and MIRI band\footnote{DJA full reduced mosaics in each band are taken from \url{https://s3.amazonaws.com/grizli-v2/JwstMosaics/v7/index.html}}, all with a pixel scale of 0.04$^{\prime \prime}$ and in units of [10 nJy]. We did a resampling of all mosaics to 0.06$^{\prime \prime}$/px for consistency with SMILES and PRIMER-COSMOS maps, while preserving astrometry and total flux over a given aperture. For PRIMER-UDS, we also merged the east and west mosaics (listed separately in the DJA) into a full mosaic.

We only considered sources whose position in the map is at least 20 pixels inside the edge of the mosaic. For each object and in each band, we created a $N$$\times$$N$ pixel image (with $N$=51, 0.06$^{\prime \prime}$/px) from the corresponding map, each centred on the NIRCam position of the LRD. We then stacked all $N$$_{\rm stack}$ sources and retrieved the median flux at each pixel.

For ALMA, instead, we performed mean stacking in the $uv$-plane, using all publicly available Band-6 observations covering our LRDs in the COSMOS and GOODS-S fields. Mean stacking is needed in order to homogeneously combine archival data at a different frequency, angular resolution, and sensitivity. This analysis is carried out with {\sc CASA}, following the method described in \citet{Wang+2022} \citep[see also][]{Wang+2024,Magnelli+2024}, to which we refer for further details of the procedure.

\subsection{Stacked fluxes}

In Fig.~\ref{fig:cutouts} we show the stacked cutouts (3$^{\prime \prime}$$\times$3$^{\prime \prime}$) in each band, sorted by increasing wavelength. For comparison, the dashed white circle marks the corresponding PSF FWHM. As MIRI images are strongly oversampled, to aid visualization we smoothed them with a Gaussian kernel, adopting a larger radius for larger PSF FWHM (see caption of Fig.~\ref{fig:cutouts}). The stacked ALMA/B6 image was obtained in the $uv$-plane and imaged at 0.1$^{\prime \prime}$/px scale. Except for MIRI/f2550w and ALMA/B6, all other median stacks lead to a detection with a signal-to-noise ratio (S/N) greater than 3. We stress that all detections are consistent with being unresolved, as their extent is smaller or comparable with the PSF FWHM of the corresponding band (white circles in Fig.~\ref{fig:cutouts}).

The uncertainty on the stacked flux density was quantified through a bootstrapping technique: if $N$$_{\rm stack}$ is the number of input targets in a given band, in each random realization we reshuffled the input sample, preserving the same $N$$_{\rm stack}$ by allowing source duplication. We repeated this 100 times and took the median of the resulting flux distribution as our formal stacked flux. The 1$\sigma$ dispersion around this value is interpreted as the flux error. The rms of the stacked image was computed by stacking the inverse-variance weighted noise map of each target. Our median stacks yield a detection in each band up to 21~$\mu$m with signal-to-noise\footnote{We consider as ``noise'' the nominal rms of the stacked image; though using the flux error derived via bootstrapping does not change our conclusions.} S/N$>$3. All fluxes, errors, and rms of the stacked images are listed in Table~\ref{tab:stacks} along with the effective number of sources stacked in each band.

We refer the reader to Appendix~\ref{Appendix_A}, where we test the robustness of our stacking results against possible biases. In particular, it is worth noting that (rms-weighted) mean stacking delivers systematically higher fluxes than median stacking, by a factor of $\times$2--3 in all bands. In Appendix~\ref{appendix_2}, we demonstrate that, at least in MIRI bands, these fluctuations are entirely due to individual MIRI detections (dropping from 76\% at f770w to $\approx$0\% at f2100w), without which the mean and median stacked fluxes are matched one another. Therefore, we reiterate that mean stacking is highly sensitive to a few brighter detections, whereas median stacking is more representative of the bulk LRD population.

\begin{table}[!t]
\caption{Median stacked fluxes and errors (open circles in Fig.~\ref{fig:sed}). }
\centering
\begin{tabular}{cc|ccc}
\hline\hline
wavelength &  $N$$_{\rm stack}$ &  rms    &  Flux &  Error Flux \\
 ($\mu$m) &                    &  (mJy)    &  (mJy)    &  (mJy)    \\
\hline
0.90 & 302 & 3.47e-09 & 8.49e-06 & 9.31e-07 \\
1.15 & 302 & 4.64e-09 & 1.31e-05 & 1.06e-06 \\
1.50 & 302 & 8.55e-09 & 1.47e-05 & 1.07e-06 \\
2.00 & 302 & 1.32e-08 & 1.78e-05 & 1.39e-06 \\
2.77 & 302 & 1.10e-08 & 3.03e-05 & 2.13e-06 \\
3.56 & 302 & 1.87e-08 & 5.89e-05 & 4.41e-06 \\
4.44 & 302 & 3.84e-08 & 1.13e-04 & 7.17e-06 \\
5.60 & 28 & 3.20e-06 & 1.23e-04 & 3.06e-05 \\
7.70 & 163 & 2.07e-06 & 2.13e-04 & 1.60e-05 \\
10.00 & 61 & 1.07e-05 & 1.94e-04 & 3.35e-05 \\
12.80 & 24 & 3.93e-05 & 2.68e-04 & 5.67e-05 \\
15.00 & 61 & 3.35e-05 & 3.42e-04 & 5.73e-05 \\
18.00 & 126 & 4.99e-05 & 5.33e-04 & 9.97e-05 \\
21.00 & 60 & 1.30e-04 & 6.08e-04 & 1.36e-04 \\
25.50 & 23 & 1.69e-03 & <5.07e-03 & --- \\
1080.00 & 70 & 5.89e-03 & <1.77e-02 & --- \\
\hline
\hline
\end{tabular} \label{tab:stacks}
\tablefoot{For each filter, the table reports the number of stacked sources ($N_{\rm stack}$), the rms of the stacked image, the median stacked flux, and its 1$\sigma$ uncertainty from bootstrapping, respectively. In the case of non-detection, the 3$\sigma$ upper limit is provided. }
\end{table}

\section{Discussion and conclusions}

\begin{figure}
\centering
     \includegraphics[width=\linewidth]{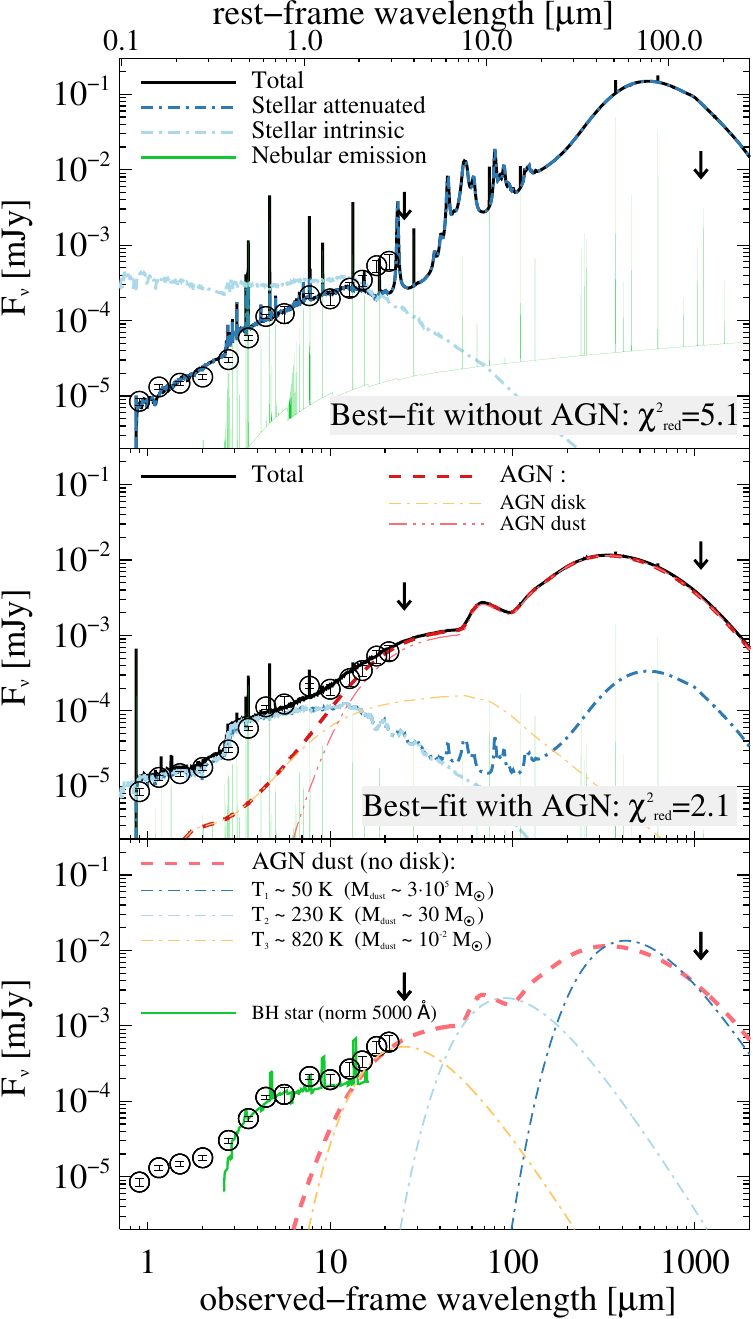}
 \caption{\small Best SED fits obtained with {\sc CIGALE} on the stacked photometry (open circles) at $z$=6.2. Top panel: Fit including only galaxy templates. Middle panel: Fit including both galaxy and AGN templates. The fitting from the AGN run is strongly favored based on $\chi^2$ arguments and ALMA constraints. Bottom panel: Decomposition of the AGN-dust template (dashed red) in three multi-temperature graybody models (dot-dashed lines). The normalized BH-star SED \citep{Naidu+2025} is overlaid for comparison (green) and not included in the SED fitting. Details are given in Section~\ref{sedfitting}.
 }
   \label{fig:sed}
\end{figure}

\subsection{SED fitting with AGN versus without AGN} \label{sedfitting}

In order to interpret the stacked photometry of LRDs, we performed SED fitting using the Code Investigating GALaxy Emission ({\sc CIGALE}; \citealt{Boquien+2019}). We fixed the input redshift to the median value of the LRD sample ($\langle{z}\rangle$$\sim$6.2). In order to account for the broad redshift distribution of the sample (mostly at 4.5$<$$z$$<$8; see Fig.~\ref{fig:z}) and the correspondingly wide rest-frame range sampled by each filter, we convolved the filter response curves at $\langle{z}\rangle$$\sim$6.2 with the band-specific redshift distribution of the underlying LRD sample. Then, we ran {\sc CIGALE} both with and without AGN templates to test the relative significance of the AGN component. In both SED runs, the library of stellar templates was taken from \citet{Bruzual+2003}, including a delayed star formation history with random bursts. Energy balance is assumed between the dust-absorbed UV radiation and the reprocessed IR emission, modeled via a dust attenuation curve from \citet{Charlot+2000}, and dust templates from \citet{Dale+2014}. The AGN templates come from the {\sc skirtor} module (\citealt{Stalevski+2012}; \citeyear{Stalevski+2016}), which includes emission from an accretion disk, a dusty torus, and polar dust. A comprehensive list of the SED-fitting parameters is given in Appendix~\ref{Appendix_C} and \ref{Appendix_D}.

In Fig.~\ref{fig:sed}, we display all stacked fluxes (or 3$\sigma$ upper limits) overlaid to the total best-fit SED (black line) obtained from the run without AGN (top panel) and with AGN (middle panel). In both runs, the total best fit (solid black line) is obtained by adding up the attenuated stellar template (dot-dashed blue line), the nebular emission (green lines), and the AGN (dashed red lines). The AGN template shown in Fig.~\ref{fig:sed} breaks down into an accretion disk component (dot-dashed yellow line) peaking in the optical NIR, and a dusty component  (clumpy torus and polar dust; dot-dashed red line) peaking at longer wavelengths. We further note that the best-fit stellar template chosen in the AGN run has very little dust attenuation (A$_V$$\sim$0.01) and the far-IR emission is therefore much fainter than the ALMA upper limit and entirely attributed to AGN (polar) dust.

Regardless of the origin of  optical and/or UV emission in LRDs (see Sect.~\ref{sed_caveats} for more details), we argue that AGN-heated dust is needed to reproduce the monothonic increase in the rest 1--3$\mu$m, which pure dust-attenuated stellar light fails to match. This is further supported by our ALMA upper limit, which is very similar to that derived by \citet{Casey+2025} (see also \citealt{Xiao+2025}; \citealt{Setton+2025}), and that translates into M$_{\rm dust}$$\lesssim$10$^6$~M$_{\odot}$ for T$\sim$100~K. Indeed, the best-fit SED without AGN strongly overshoots the ALMA upper limit when attempting to reproduce the power-law-like SED at rest 1--3$\mu$m. While this could be resolved by lifting the energy balance implemented in {\sc CIGALE}, even ignoring the upper limits (at f2550w and ALMA/B6), a simple F-test comparing the two reduced $\chi^2$ values favors the AGN solution at $>$95\% significance.

\subsubsection{Caveats and limitations of broadband SED fitting} \label{sed_caveats}

We caution the reader that our SED analysis does not necessarily endorse a pure stellar origin of the optical and/or UV continuum, despite the stellar template chosen by {\sc CIGALE} in the SED fitting. We highlight some caveats and limitations inherent to our photometric SED analysis. For instance, {\sc CIGALE} assumes a single-zone obscuration between optical, UV, and IR light, which might be oversimplistic in the case of no co-spatiality (see Sect.~\ref{discussion}). In addition, broad emission lines - such as those observed in most LRD spectra - are not properly plugged into the CIGALE library. More importantly, despite the fact that our median stacking mitigates broad emission line contamination, it relies on broadband photometry, which most likely irons out the strongest Balmer breaks, with a potential impact on the output stellar ages and $M_{\star}$.

Regarding limitations on the AGN templates, instead, while a BH-Star model \citep{Naidu+2025}, or a quasi-star model \citep{Santarelli+2025}, could be better suited for reproducing the strong Balmer breaks, Balmer line absorption, and broad emission lines (e.g., \citealt{deGraaff+2025}), a satisfactory fit of the full SED (as shown in Fig.~\ref{fig:sed}, bottom panel) further requires a dusty torus and (in most cases) an additional UV component. In this framework, accommodating a joint BH star (or quasi star) and a dusty-torus template demands radiative transfer modeling (following e.g., {\sc skirtor}; \citealt{Stalevski+2012}; \citeyear{Stalevski+2016}).

Therefore, a more faithful modeling of the spectro-photometric LRD features would likely require an ad-hoc grid of (AGN) models with those built-in features. Besides the uncertainty related to the representativeness of such realistic models, this task is beyond the scope of this paper, which focuses on hot-dust emission from stacking photometric data.

\subsection{Comparison with literature LRD SEDs} \label{sed_comparison}

\begin{figure}
\centering
     \includegraphics[width=\linewidth]{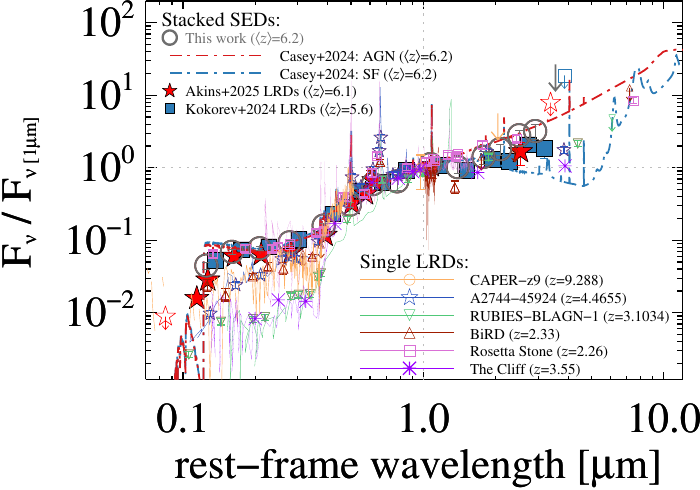}
 \caption{\small {Comparison between our rest-frame median stacked SED (open gray circles) and other literature spectra or SEDs of LRDs. All fluxes are normalized to the rest-frame 1~$\mu$m flux density. We show the stacked SEDs from \citet{Kokorev+2024}; \citet{Akins+2025} and \citeauthor{Casey+2024} (\citeyear{Casey+2024}; both AGN and galaxy fits as red and blue dot-dashed lines, respectively). Other single LRD SEDs include: CAPER-z9 ($z$=9.288; \citealt{Taylor+2025}); A2744-45924 ($z$=4.4655; \citealt{Labbe+2024}); RUBIES-BLAGN-1 ($z$=3.1034; \citealt{Wang+2025}); the BiRD ($z$=2.33; \citealt{Loiacono+2025}); the Rosetta Stone (or GN 280754; $z$=2.26; \citealt{Juodzbalis+2024a}) and The Cliff ($z$=3.55; \citealt{deGraaff+2025}).}}
   \label{fig:sed_literature}
\end{figure}

In the literature, LRDs are inevitably selected with distinct criteria, depending on the availability of multi-band NIRCam and MIRI imaging, and rest-frame optical spectra. While our LRD sample meets the same compactness and V-shape criteria across all fields, we compare our stacked SED against other LRD SEDs based on different selections. {Fig.~\ref{fig:sed_literature} displays a number of stacked and single LRD SEDs for comparison, all in the rest frame and normalized to the rest 1~$\mu$m flux density to emphasize the difference in NIR slope}.

Among the median stacked SEDs, we show: \citeauthor{Akins+2025} (\citeyear{Akins+2025}; red starred symbols); the best-fit galaxy and AGN SED fits from \citeauthor{Casey+2024} (\citeyear{Casey+2024}; dot-dashed lines); and the median stacked SED obtained from the LRD sample of \citeauthor{Kokorev+2024} (\citeyear{Kokorev+2024}; blue squares). \citet{Akins+2025} selected 434 LRDs in the COSMOS-Web area (0.5~deg$^2$; \citealt{Casey+2023}) that simultaneously meet a compactness cut (in f444w), a very red color cut (f277w-f444w$>$1.5), and a bad brown dwarf fit, without explicitly requiring a V-shaped SED due to limitations in NIRCam imaging. Their median SED agrees remarkably well with ours, though their statistics in MIRI is limited to 7.7$\mu$m (over 1/3 of COSMOS-Web) and 18$\mu$m (from the inner PRIMER-COSMOS footprint, $\approx$4\% of the full sample), this latter leading to tentative detection (0.43$\pm$0.15~mJy). We show that their 18$\mu$m flux is fully consistent with the rising MIR slope seen longward 10$\mu$m. Their upper limits are here re-scaled to 3$\sigma$. Despite the different LRD selection and survey field ($\lesssim$25\% of our LRDs are in \citealt{Akins+2025}), this agreement further suggests that the compactness and the conservative color selection by \citet{Akins+2025} broadly compensate for the need for a V-shaped SED, which leads to self-consistent LRD samples.

\citet{Kokorev+2024} selected 260 LRDs over the same JWST Legacy fields as this study, based on a compactness (in f444w) criterion and various color cuts that mimic a V-shape requirement (see Sect.3.1 in \citealt{Kokorev+2024} for details). From their original sample, we further updated the redshift measurements with additional $z_{\rm spec}$ as done in Sect.~\ref{sample}, and repeated the same median stacking analysis carried out for our LRD sample; we found good consistency.

In Fig.~\ref{fig:sed_literature} we also compare our stacks against the median best-fit SED by \citet{Casey+2024}, which was obtained from stacking a joint LRD sample (675 LRDs) from \citet{Akins+2025} and \citet{Kokorev+2024}. In their work, \citet{Casey+2024} performed stacking in rest-frame wavelength bins, applying a K-correction based on the observed SED of each LRD. The median stacked SED was then redshifted to $z$=6.2 (as for our sample) and fitted with either a pure AGN (dashed red) or a pure galaxy template (dashed blue). We emphasize that our median stacks are fully consistent with their median AGN SED within their uncertainties (not reported here). Our ALMA upper limit at $\approx$1.1~mm is also fully consistent with the ALMA~1.3~mm upper limit derived in \citet{Casey+2025} via stacking 60 LRDs retrieved from the \citet{Akins+2025} sample.

In addition, in Fig.~\ref{fig:sed_literature} we display the spectra and photometric SEDs of some well-studied LRDs for comparison, namely: CAPER-z9 ($z$=9.288; \citealt{Taylor+2025}); A2744-45924 ($z$=4.4655; \citealt{Labbe+2024}); RUBIES-BLAGN-1 ($z$=3.1034; \citealt{Wang+2025}); the ``Big Red Dot'' (BiRD, $z$=2.33; \citealt{Loiacono+2025}); the ``Rosetta Stone'' (or ``GN 280754''; $z$=2.26; \citealt{Juodzbalis+2024a}) and ``The Cliff'' ($z$=3.55; \citealt{deGraaff+2025}). This compilation showcases the variety of rest-frame 1--3~$\mu$m SEDs of bright LRDs, as compared to median stacked SEDs. As noted here and in other works (e.g., \citealt{Barro+2024}; \citealt{Loiacono+2025}; \citealt{Setton+2025}; \citealt{Ronayne+2025}), most of the brightest LRDs do not seem to show a rising NIR slope but flat, or even decreasing, rest-frame 1--3~$\mu$m SEDs. This is at odds with recent MIRI stacked SEDs, which are dominated by non-detections (only $\lesssim$10\% are detected longward f1000w, see Appendix~\ref{appendix_2}). Such a comparison further highlights that LRDs are not a single homogeneous population, and that our median stacking in the longest MIRI bands is more representative of intrinsically fainter sources.

Another meaningful comparison concerns the work by \citet{Williams+2024}, who stacked LRDs over all MIRI bands within the GOODS-SMILES footprint. They found a flat NIR SED up to 18$\mu$m (and upper limits longward), thus excluding a dominant contribution from AGN-heated dust. Besides the small numbers (only nine objects) and the slightly higher redshift compared to our sample ($\langle{z}\rangle$$\sim$7.7 instead of $\langle{z}\rangle$$\sim$6.2), their LRDs were pre-selected from a pool of H-band dropouts (i.e., Hubble Space Telescope; HST-dark galaxies), hence likely biased toward the most obscured sources. We repeated the stacking for the same nine sources and found the same median SED as in \citet{Williams+2024}. This check suggests that their sample might be sensitive to more obscured sources than conventional LRDs, especially considering their H-dropout selection, combined with the lack of compactness and V-shaped requirements. These differences might link to a different obscuration origin relative to typical LRDs. Hence we caution that these nine LRDs are not representative of our photometrically selected LRDs (from \citealt{Kocevski+2025}), despite being an interesting subsample to follow up.

\subsection{Are LRDs typically dust-obscured AGN?} \label{discussion}

By considering the best-fit solution with AGN, the rest-frame 6~$\mu$m luminosity of the AGN torus can be used as a proxy for the intrinsic AGN X-ray (rest [2--10]~keV) luminosity ($L$$_{\rm X, intr}$ (e.g. \citealt{Stern+2015}). From $L$$_{\rm 6\mu m}$=(3.8$\pm$0.6)$\times$10$^{43}$~erg/s, we derive $L$$_{\rm X, intr}$=2$\times$10$^{43}$~erg/s, which nicely matches that expected from the best-fit AGN bolometric luminosity, if we assume standard X-ray bolometric corrections (e.g., \citealt{Lusso+2012}). By stacking the deepest X-ray images of the 44~LRDs that fall inside the Chandra Deep Field South \citep{Luo+2017}, we obtain a 3$\sigma$ [2--10] keV luminosity limit $L$$_{\rm X}$$\lesssim$10$^{42}$~erg/s at $z$=6.2 \citep{Comastri+2025}. In order to match it with the $L$$_{\rm X, intr}$ expected from the AGN SED, the hydrogen column density has to be N$_{\rm H}$$\gtrsim$3$\times$10$^{24}$~cm$^{-2}$, which implies Compton-thick gas obscuration (\citealt{Ananna+2024}; \citealt{Yue+2024}; \citealt{Maiolino+2025}; \citealt{Sacchi+2025}).

Given the SED-based estimate of the AGN bolometric luminosity, $L_{\rm bol,AGN} = (2.5 \pm 0.4) \times 10^{44}$~erg~s$^{-1}$ and the best-fit stellar mass from the AGN run (M$_{\star}$$\sim$2.5$\times$10$^9$~M$_{\odot}$; i.e., $\approx$2$\times$ smaller than from a pure-galaxy fit), we emphasize that, if LRDs are highly accreting AGN (Eddington ratio $\gtrsim$10\%), the BH-to-galaxy mass ratio M$_{\rm BH}$/M$_{\star}$$\gtrsim$1:100, implies strongly overmassive BHs compared to local scaling relations (e.g. \citealt{Reines+2015}). Moreover, our $L_{\rm bol,AGN}$ estimate corresponds to a dust sublimation radius of $R_{\rm sub}$$\simeq$0.1~pc at $T_{\rm sub}$$\simeq$$1500$~K \citep{Barvainis1987}, meaning that any dusty AGN structure should begin at scales $\gtrsim 0.1$~pc from the central SMBH. In the bottom panel of Fig.~\ref{fig:sed}, we try to decompose the AGN-dust template (dashed red curve) with three multi-temperature graybody curves (leaving $T$ free to vary). Our rising MIRI SED up to rest-frame 2--3$\mu$m indicates hot dust at $T$$\simeq$$820$~K, though it could be even hotter if optically thick. Instead, the bulk dust mass (M$_{\rm dust}$$\approx$3$\times$10$^5$~M$_{\odot}$) comes from colder polar AGN dust at $\approx$50~K (though not constrained by our data). While a dusty structure could induce X-ray obscuration, it would also strongly suppress the optical and/or UV continuum, in tension with the blue UV slope and strong Balmer break features that imply relatively modest dust attenuation ($A_V$$\sim$$2{-}3$; \citealt{Akins+2025}; \citealt{Casey+2025}).

This apparent contradiction can be resolved if X-ray absorption in LRDs occurs within $R_{\rm sub}$ ($<$0.1~pc), in a gas-dense and dust-free region, implying no co-spatiality. We stress that, despite the fact that our stacking analysis suggests that most ($\gtrsim$50\%) LRDs contain hot-AGN dust, they do not need to be dust obscured. This gas-dust displacement implies that theoretical models of quasi stars \citep{Begelman+2025}, BH stars \citep{Naidu+2025}, and super-massive stars \citep{Nandal+2025}, can in principle co-exist with outer hot dust, albeit its formation scenario (in-situ or ex-situ) remains open.

Our findings also fit with an alternative hypothesis (\citealt{Setton+2024}; \citealt{Maiolino+2025}; \citealt{DEugenio+2025a}; \citealt{Ji+2025}; \citealt{Inayoshi+2025}) that postulates that X-ray obscuration arises from Compton-thick gas clouds within the BLR (at $R$$<$$R$$_{\rm sub}$, i.e., dust-free), which covers the compact X-ray corona without suppressing the broad Balmer lines. However, the origin of the optical and/or UV continuum remains open. In this debated framework, recent works report the presence of a forest of auroral and FeII lines in some luminous LRDs (e.g., \citealt{Tripodi+2025}; \citealt{Lin+2025}; \citealt{Lambrides+2025}; \citealt{Torralba+2025}; \citealt{DEugenio+2025b}), which indicate that - at least in these bright objects - part of the UV continuum is dominated by a compact AGN embedded in a dense gas cocoon that (at least partly) explains the Balmer break and the broadening of the Balmer lines. Regardless of the AGN orientation, we argue that the hot dust revealed in this work should be placed most likely at outer radii (R$_{\rm sub}$$>$0.1~pc) than the BLR scales at which the (dust-free) gas absorption takes place. Given that R$_{\rm sub}$$\propto$$L_{\rm bol,AGN}^{1/2}$, we expect that even in the most luminous LRDs (e.g., \citealt{Torralba+2025}; \citealt{DEugenio+2025b}, i.e., 50-100$\times$ brighter than our median stacked AGN), R$_{\rm sub}$$\gtrsim$1~pc, which is $\approx$10$^4$ times larger than the nominal thickness of the BLR (gas) cloud \citep{DEugenio+2025b}.

\begin{figure}
\centering
     \includegraphics[width=\linewidth]{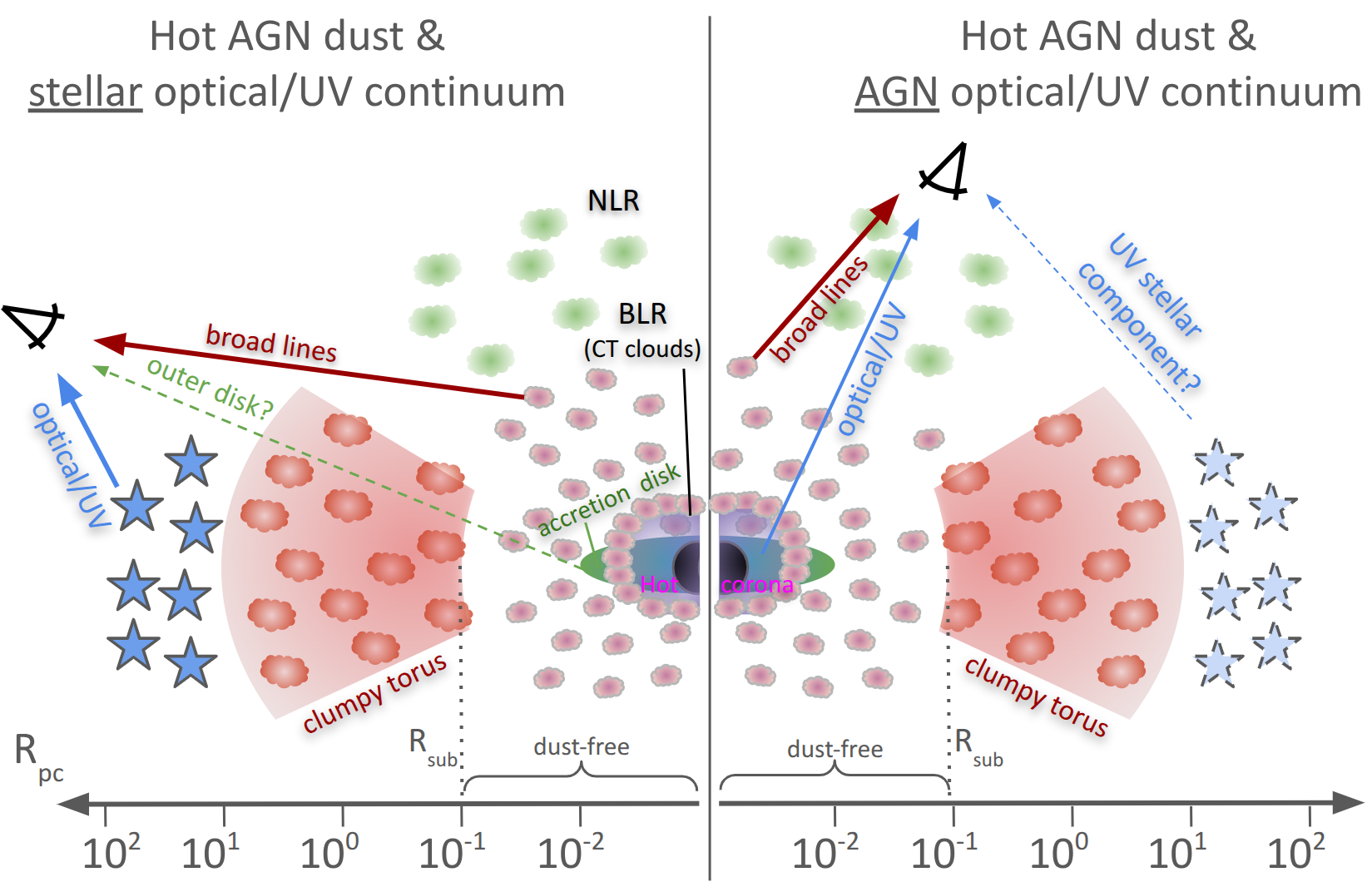}
 \caption{\small Sketch of the proposed hot AGN-dust scenario, both for stellar-dominated (left) and AGN-dominated (right) optical and/or UV continuum.
 }
   \label{fig:sketch}
\end{figure}

Although we caution that LRDs might not consist of a single population, we speculate two simple scenarios in Fig.~\ref{fig:sketch}, depending on what powers the optical and/or UV continuum in LRDs. In either case, the observed optical and/or UV continuum comes from dust-unattenuated regions, despite the presence of hot AGN dust, and the outer BLR is directly visible along the line of sight. \\
(i) If the optical and/or UV comes from unobscured stellar light (Fig.~\ref{fig:sketch}, left), it must be placed at larger scales than the AGN torus. In this case the LRD resembles a Type-2 AGN configuration, in which the torus covers the AGN accretion disk, which would otherwise outshine the host. In the case of high torus clumpiness, mildly attenuated light from the outer accretion disk and the BLR could pierce through the torus and add to the optical and/or UV emission. \\
(ii) Instead, if the optical and/or UV is dominated by AGN light (Fig.~\ref{fig:sketch}, right), the LRD resembles a Type-1 unobscured AGN configuration, though the origin of the sharp Balmer break remains unclear. Fainter optical and/or UV light arising from (kiloparsec-scale) star formation cannot be ruled out (e.g., \citealt{Rinaldi+2025}).

\section{Conclusions}

In this manuscript, we investigate the nature of LRDs via an extensive NIRCam, MIRI, and ALMA stacking analysis of a homogeneously selected sample of 302 sources across the CEERS, JADES, and PRIMER JWST fields. We fit the median stacked SED with either pure galaxy or galaxy+AGN templates to interpret the average IR emission properties of LRDs.

In summary, we find clear evidence for a rising MIRI continuum up to $\lambda_{\rm rest}$$\sim$3$\mu$m, which we interpret via SED fitting as hot ($\simeq$820~K) AGN dust, regardless of whether the optical and/or UV continuum is stellar- or AGN-dominated, or a combination thereof. Despite the presence of hot dust, these LRDs selected via V-shaped and compactness criteria must be not strongly dust-obscured. The non-detection from deep X-ray stacking suggests that the apparent X-ray weakness of LRDs is caused by Compton-thick ($N_{\rm H}$$>$3$\times$10$^{24}$~cm$^{-2}$) gas obscuration.

We acknowledge that our homogeneous LRD selection does not necessarily translate into a single LRD population with similar physical properties. However, we reiterate that our main empirical finding is unchanged, that is, the majority ($\gtrsim 50\%$) of LRDs show AGN-heated dust emission. If this were not the case, the rising NIR slope would not appear in the median stacked data, and the mean and median MIRI fluxes would differ significantly after removing individual detections (contrary to our findings in Appendix~\ref{appendix_2}). The small fraction of individual MIRI detections ($\lesssim$10\% longward f1000w, i.e., longer than rest 1.3$\mu$m) contributes negligibly to the median stacked SED at rest-frame 1--3$\mu$m. Therefore, the rising NIR slope is predominantly driven by the MIRI non-detections, which suggests that AGN-heated dust is characteristic of the fainter, more typical LRD population.

That being said, our results are contingent upon the adopted LRD selection criteria, which may be inherently biased. For instance, if the rest-frame UV emission of LRDs arises from an AGN accretion disk, we speculate that photometric LRD selection criteria (especially the blue UV-slope requirement) would preferentially select Type-1 AGN, according to the Unified Model. The same criterion would therefore induce a bias against the Type-2 LRD analogs, in which the disk is hidden by an obscuring dusty structure. With this caveat in mind, our results should be taken as an empirical benchmark for motivating deeper investigations of the full LRD population.

\begin{acknowledgements}
The authors thank the anonymous referee for their constructive report. ID thanks Federica Loiacono for sharing their SED datapoints, and also Giovanni Mazzolari, Yoshiki Matsuoka and Guillermo Barro for insightful discussions. ID acknowledges funding by the European Union -- NextGenerationEU, RRF M4C2 1.1, Project 2022JZJBHM: "AGN-sCAN: zooming-in on the AGN-galaxy connection since the cosmic noon" - CUP C53D23001120006. SB is funded by ERC grant 101076080. MG wishes to acknowledge the generous support and hospitality, by the CEA -  Universit\'e Paris-Saclay during his visit, when part of the research related to this work was carried out. (Some of) the data producs presented here were retrieved from the DJA, an initiative of the Cosmic Dawn Center (DAWN), which is funded by the Danish National Research Foundation under grant DNRF140.
We acknowledge the contribution of the COSMOS collaboration, consisting of more than 200 scientists. More information about the COSMOS survey can be found at \url{https://cosmos.astro.caltech.edu/}. This work is based [in part] on observations made with the NASA/ESA/CSA James Webb Space Telescope. The data were obtained from the Mikulski Archive for Space Telescopes at the Space Telescope Science Institute, which is operated by the Association of Universities for Research in Astronomy, Inc., under NASA contract NAS 5-03127 for JWST. These observations are associated with program $\#$1180, $\#$1181, $\#$1210, $\#$1286, $\#$1895, $\#$1963, $\#$3215, $\#$1345, $\#$1837 and $\#$1207. The data described here may be obtained from the MAST archive at \url{https://doi.org/10.17909/et3f-zd57} (SMILES) and \url{https://dx.doi.org/10.17909/8tdj-8n28} (JADES) and on the DJA at \url{https://s3.amazonaws.com/grizli-v2/JwstMosaics/v7/index.html} (CEERS and PRIMER-UDS).
\end{acknowledgements}

 \bibliographystyle{aa}
 \bibliography{biblio}

\appendix

\section{Stacking tests} \label{Appendix_A}

\begin{figure}
\centering
     \includegraphics[width=\linewidth]{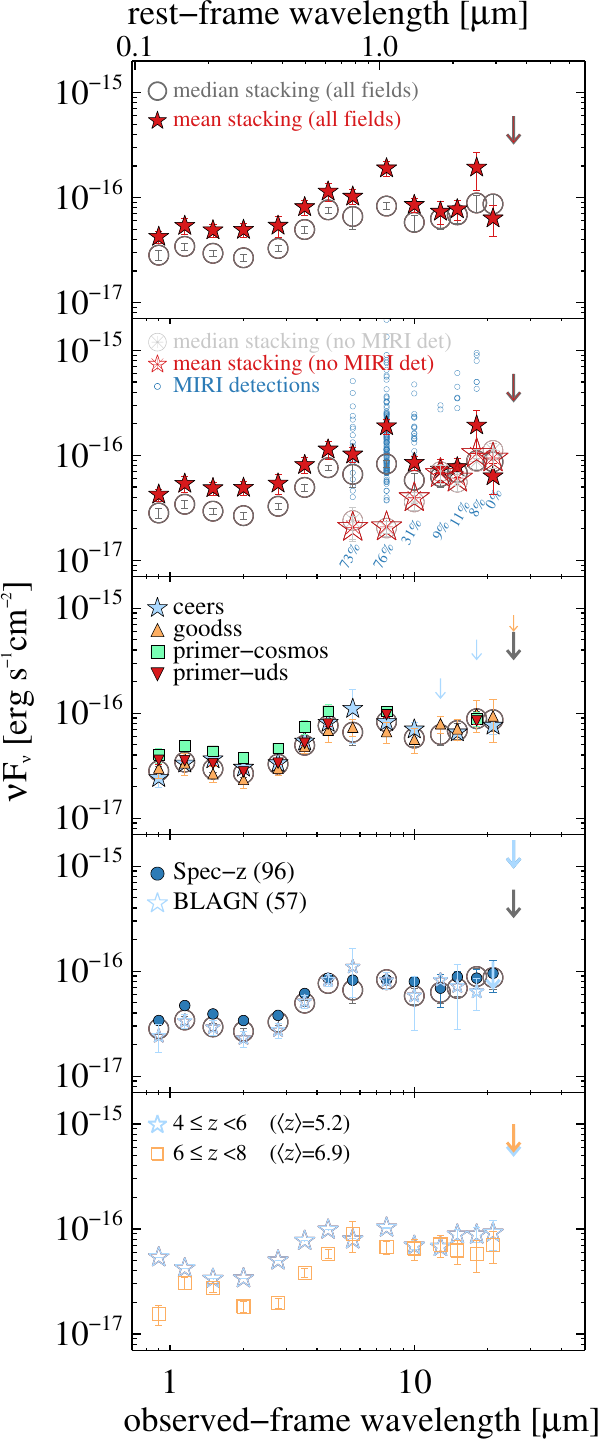}
 \caption{\small Summary plot comparing our median stacked fluxes (grey open circles) against multiple effects. From top to bottom: mean vs median stacking (Appendix~\ref{appendix_1}); contribution of MIRI detections (Appendix~\ref{appendix_2}); field-to-field variations (Appendix~\ref{appendix_3}); effect of photometric redshifts (Appendix~\ref{appendix_4}); effect of a broad redshift range (Appendix~\ref{appendix_5}). Details are given in the corresponding sub-sections.
 }
   \label{fig:appendix}
\end{figure}

We test our stacking results against multiple potential biases related to sample selection, stacking method, field-to-field variations and photometric redshift uncertainties. The main outcome of these tests is shown in Fig.~\ref{fig:appendix}. In general, these tests further corroborate the robustness of our stacking results and the need for an obscured AGN component to reproduce the MIR excess. Each of the following tests is discussed in a dedicated subsection, numbered as the panels of Fig.~\ref{fig:appendix} from top to bottom.

\subsection{Mean vs median stacking} \label{appendix_1}

In the first panel of Fig.~\ref{fig:appendix} we compare our median stacked fluxes (grey open circles) against mean stacked fluxes (red starred symbols). These latter values are computed from the inverse-variance weighted flux density map, accounting for the different sensitivity (i.e. exposure time) of each target. Mean stacking leads to higher fluxes with respect to median stacking. This is likely caused by the fact that the resulting (rms-weighted) linear mean flux can be boosted upwards by relatively bright individual detections (we note that upper limits at f2550w and ALMA/B6 are unchanged as they are calculated from the rms map). This is the case in NIRCam bands, where all LRDs are individually detected by definition over multiple bands, both blueward and redward the Balmer break (in order to assess the V-shape). An even stronger boost is seen at MIRI 7.7$\mu$m and 18$\mu$m, which count about 100 more targets than at other MIRI bands (see Table \ref{tab:sample} and Appendix~\ref{appendix_2}). In these two MIRI bands, we investigate the origin of these fluctuations, finding that the f770w and f1800w fluxes are over-boosted relative to other bands because of two strong outliers at $z$$<$4.5 in the PRIMER-UDS field. These are: RUBIES-BLAGN1 (or PRIMER-UDS 38223, at $z$=$_{\rm spec}$=3.1; \citealt{Wang+2025}), which is a bright X-ray detected BLAGN; and PRIMER-UDS 9235 ($z$=$_{\rm phot}$=4.45), previously flagged in \citet{Leung+2025} to be the strongest f1800w emitter of the PRIMER-UDS field. We check that after removing only those two PRIMER-UDS outliers, the ratio between mean and median stacked fluxes (over all fields) re-aligns to the average factor of $\times$2-3 seen in all other bands. We do not remove these outliers in the rest of our work, in order to not fine-tune the results. However, this is again a caveat on using mean stacks without pre-removing bright individual detections. Moreover, we opt for median stacking also because it strongly mitigates contamination from bright neighbors, and thus reduces the confusion noise for the faint sources. This comparison, however, demonstrates that mean and median stacked SEDs display broadly-similar shapes.

\subsection{Contribution of MIRI detections to the stacking} \label{appendix_2}

We test the robustness of our stacking results against the contribution of individual detections. We repeat both mean (open stars) and median (open circles) stacking after removing individual MIRI detections (with S/N$>$3), in each band. In Figure~\ref{fig:appendix} (2$^{\rm nd}$ panel from top) we display the new stacking results. The individual MIRI detections are shown as small blue circles, and the corresponding detection rate is reported at each MIRI band. We notice that mean and median stacking of MIRI non-detections lead to well-consistent results, also in f560w and f770w where the detection rate is above 70\%. In addition, for MIRI non-detections we still observe a significant rise at rest-frame 1--3~$\mu$m, while individual MIRI detections stand out at much higher fluxes. This check strongly suggests that the difference between median and mean stacks boils down to the number of individual detections, and that the rising near-IR slope is driven by the bulk of non-detections, which is sensitive to more common and intrinsically fainter sources.

\subsection{Field-to-field variations} \label{appendix_3}

An important aspect that our analysis relies on is the combination of multiple NIRCam and MIRI survey fields. While this is necessary to improve the statistics and the stacked S/N, it might potentially introduce systematics related to different JWST coverage and depth, which could affect the properties of photometrically-selected LRDs. We thus verify that our stacking results are robust against field-to-field variations, by repeating the same analysis separately for each field. The third panel in Fig.~\ref{fig:appendix} displays the median stacked SED obtained in each field (represented by different filled symbols) overlaid to our full stacking results (grey open circles). For visual purposes, here we relax the detection threshold to S/N$>$2 (hence upper limit fluxes are set to twice the noise level if S/N$<$2). Despite their heterogeneous spectral coverage across MIRI bands, all single-field median SEDs are consistent with the combined median SED. {Moreover, we verify that stacking the same exact 22 objects having simultaneously NIRCam and MIRI coverage in all bands (from the GOODS-S field), the resulting SED is fully consistent with that obtained via stacking the full LRD sample. These checks strengthen that similarly-selected LRDs across different fields share a common median SED, despite their and heterogeneous NIRCam/MIRI spectral coverage.}

\subsection{Effect of photometric redshifts} \label{appendix_4}

A relevant caveat in our analysis is the photometric selection of LRDs. Specifically, photometric redshift uncertainties can be notably high, potentially biasing the LRD selection. We iterate that for the spectroscopic subsample (96/302 objects) there is good agreement between photometric and spectroscopic redshifts (Sect.~\ref{sample}). However, we test this directly by stacking NIRCam and MIRI images solely for the spectroscopic subsample. The fourth panel in Fig.~\ref{fig:appendix} displays the corresponding median stacked SED (blue filled circles). Numbers are given in Table~\ref{tab:sample} for each field: when combining all fields there are between 36 and 53 objects in MIRI (except for f560w, f1280w and f2550w with $\leq$5 objects due to a smaller areal coverage) and 96 objects in NIRCam. Despite the S/N being smaller than that of the full sample, the two median stacked SEDs agree remarkably well with each other.

We also explore whether a cleaner selection of ``bona fide'' AGN based on broad emission lines yields similar trends. From the NIRSpec Merged Table of the DJA, we isolate 57/96 ($\sim$59\%) objects with broad H$_{\alpha}$ or H$_{\beta}$ emission lines, as inferred from the best-fit continuum fluxes, line intensity and equivalent widths. This fraction agrees with that reported in the literature (e.g. \citealt{Greene+2024}; \citealt{Hviding+2025}). The corresponding median stacked SED of BL AGN (empty starred symbols), displays no significant difference with respect to the full spec-$z$ (96) subsample, albeit a lower S/N in all bands due to smaller statistics.

\subsection{Effect of a broad redshift range} \label{appendix_5}

As shown in Fig.~\ref{fig:z}, our targets span a very broad redshift range, from $z$$\sim$2 to $z$$\sim$11, with over 90\% of the sample being within the range (4$<$$z$$<$8). Because we perform stacking in the observed frame, the contribution to each stacked point comes from very different parts of the rest-frame spectrum. As mentioned in Sect.~\ref{sedfitting}, we already account for this by convolving the filter response with the underlying redshift distribution. However, this approach assumes that each galaxy contributes evenly to the stack, regardless of its redshift, whereas the flux dimming can vary significantly across the full range. We thus test the impact of the redshift range to the stacked fluxes by splitting our LRDs into two similarly-populated $z$-bins at 4$\leq$$z$$<$6 (126 LRDs) and 6$\leq$$z$$<$8 (121 LRDs). We show this comparison in the bottom panel of Fig.~\ref{fig:appendix}. The lower-$z$ bin shows slightly brighter fluxes and steeper MIRI slope compared to the higher-$z$ bin, as expected given the smaller distance and the longer rest-frame wavelength probed (e.g. 3.5~$\mu$m instead of 2.5~$\mu$m at MIRI-21~$\mu$m), respectively. {We also note that fluxes are in energy units, hence the apparent flattening of the 6$\leq$$z$$<$8 NIR SED translates into a monothonic rise in flux density ($F_{\nu}$) units, as seen in the full sample.} Overall, this check suggests that LRDs selected at different redshifts have broadly-consistent stacked SEDs, with no evidence for a galaxy-like drop at rest 1--2~$\mu$m. We further note that emission lines are unlikely to systematically boost the median stacks, as this would be the case if a large fraction of sources were clustered in a narrow redshift range (such that a rest-frame emission line would systematically fall into a given filter). Given the broad similarity of the two SEDs in two different $z$-bins, we believe this is unlikely the case.

\clearpage

\section{Input SED-fitting parameters} \label{Appendix_C}

   \begin{table}[!h]
   \caption{Main input parameters for SED fitting with {\sc CIGALE} (Section \ref{sedfitting}).  \label{tab:sedfitting}}
   \centering
   \small
   \begin{tabular}{clcccccc}
   \hline\hline
   Module & Parameter & Symbol & Values \\
   \hline\xrowht[()]{5pt}
                                          & Stellar $e$-folding time & $\tau_{\rm{star}}$ ($10^6$ yr) & 150, 200, 300, 500, 700, 900$^{*}$ \\
Star formation history                    & Stellar age      & $t_{\rm{star}}$ ($10^6$ yr)     & 200, 300, 500, 700, 900$^{*}$ \\
$[\rm{SFR} \propto t \exp(-t/\tau)]$      & Age of the burst                     & $t_{\rm{burst}}$ ($10^6$ yr)    & 10, 30, 50$^{*}$, 100  \\
                                          & SFR ratio after/before the burst     & $r_{\rm SFR}$                      & 0.1, 0.5, 1$^{*}$, 2, 5, 10 ,30, 50 \\
\hline\xrowht[()]{5pt}

Single stellar population      & Initial mass function      & --                                              & \citet{Chabrier2003} \\
$[$\citep{Bruzual+2003}$]$      & Metallicity                     & $Z$          &            0.0004, 0.008, 0.02$^{*}$ \\
  \hline\xrowht[()]{5pt}

\multirow{3}{*}{Nebular emission} & Ionization parameter       & $\log U$          & -2.0$^{*}$, -1.0 \\
                                  & Gas metallicity            & Z$_{\rm gas}$ & 0.0004$^{*}$, 0.008, 0.019 \\
                                  & Electron density           & $n_e$         &       100,1000$^{*}$ \\
\hline\xrowht[()]{5pt}
                                  & V-band attenuation (ISM)    & A$_{\rm V, ISM}$ & 0.01$^{*}$, 0.1, 0.3, 0.5, 0.7, 1.0, 1.5, 2, 2.5, 3, 3.5 \\
Dust attenuation                  & Fractional ISM attenuation    & A$_{\rm V, ISM}$ / (A$_{\rm V, BC}$+A$_{\rm V, ISM}$)   & 0.44$^{*}$  \\
$[$mod. \citep{Charlot+2000}$]$     & Power-law slope (ISM)    & --   & -0.7$^{*}$  \\
                                  & Power-law slope (BC)    & --   & -1.3$^{*}$ \\
\hline\xrowht[()]{5pt}
                            & Mass fraction of PAH                                & q$_{\rm pah}$     &   0.47, 2.5, 3.9$^{*}$  \\
Galactic dust emission      & Minimum radiation field                             & U$_{\rm min}$     & 1.0, 10, 20, 40$^{*}$        \\
$[$\citep{Dale+2014}$]$             & Slope in $dM_{\rm{dust}} \propto U^{-\alpha}dU$     & $\alpha$          &  2.0$^{*}$ \\
                            & Illuminated fraction [Umin--Umax]              & $\gamma$          & 0.02$^{*}$  \\
\hline\xrowht[()]{5pt}

                           &  Edge-on optical depth at 9.7$\mu$m      & $\tau_{9.7}$     &      3$^{*}$, 7        \\
                           &  Slope of radial dust density profile    &    --            &      1.0$^{*}$          \\
                           &  Slope of polar dust density profile     &    --            &      1.0$^{*}$          \\
                           &  Torus opening angle (from equator)      &   $\Theta$       &    40$^{*}$            \\
AGN disk+torus+polar dust: &  Ratio outer/inner torus radius          &   --             &     20$^{*}$            \\
{\sc skirtor}              &  Viewing angle (from vertical axis)      & $i$              &    50$^{*}$, 70            \\
$[$\citeauthor{Stalevski+2012} (\citeyear{Stalevski+2012}; \citeyear{Stalevski+2016})$]$ &  AGN fraction                          &     $f_{\rm AGN}$      &    0.0, 0.1, 0.3, 0.5, 0.7, 0.9, 0.99$^{*}$    \\
                           &  Extinction law for polar dust           &     --           &     SMC                      \\
                           &  $E(B-V)$ in the polar direction         &  --              &     0.00, 0.05$^{*}$, 0.10         \\
                           &  Temperature of polar dust               &      --          &     100~K$^{*}$                    \\
                           &  Emissivity of polar dust                &  --              &        1.6$^{*}$                   \\
\hline
   \end{tabular}
   \tablefoot{(${*}$) Asterisks mark the best-fit parameters obtained from the AGN run.}

   \end{table}

\section{Output SED-fitting parameters} \label{Appendix_D}

   \begin{table}[!h]
   \caption{List of relevant best-fit output parameters obtained with {\sc CIGALE} (Section \ref{sedfitting}). }
   \centering
   \small
   \begin{tabular}{c|c}
   \hline\hline
   Parameter & Value and $\pm$1$\sigma$ error \\
   \hline\hline
   Stellar mass (fit without AGN)   & (5.0$\pm$1.0)$\times$10$^9$~M$_{\odot}$   \\
   Stellar mass (fit with AGN)   &  (2.5$\pm$0.3)$\times$10$^9$~M$_{\odot}$  \\
   Total dust luminosity (fit without AGN)     &  (5.4$\pm$1.2)$\times$10$^{44}$~erg/s   \\
   Total dust luminosity (fit with AGN)   &  (1.4$\pm$0.2)$\times$10$^{44}$~erg/s  \\
   SFR [100~Myr] (fit without AGN) & (8.2$\pm$3.6) M$_{\odot}$/yr \\
   SFR [100~Myr] (fit with AGN) & (0.6$\pm$0.1) M$_{\odot}$/yr \\
   SFR [10~Myr] (fit without AGN) & (15.9$\pm$5.2) M$_{\odot}$/yr \\
   SFR [10~Myr] (fit with AGN) & (0.5$\pm$0.2) M$_{\odot}$/yr \\
   AGN bolometric luminosity  &  (2.5$\pm$0.4)$\times$10$^{44}$~erg/s  \\
   AGN 6~$\mu$m luminosity  &  (3.8$\pm$0.6)$\times$10$^{43}$~erg/s  \\
   \hline
   \end{tabular}
      \tablefoot{Uncertainties are reported at $\pm$1$\sigma$ and retrieved from the likelihood distribution function recorded in the fitting. }
\label{tab:sedfitting}
   \end{table}

\end{document}